# Connected Learning, Collapsed Contexts

Examining teens' sociotechnical ecosystems through the lens of digital badges

CAROLINE PITT

The Information School, University of Washington, Seattle, Washington, USA, pittc@uw.edu

ADAM BELL

College of Education, University of Washington, Seattle, Washington, USA, abell42@uw.edu

BRANDYN S. BOYD

The Information School, University of Washington, Seattle, Washington, USA, brandyn8@uw.edu

NIKKI DEMMEL

The Information School, University of Washington, Seattle, Washington, USA, ndemmel@uw.edu

KATIE DAVIS

The Information School, University of Washington, Seattle, Washington, USA, kdavis78@uw.edu

Researchers and designers have incorporated social media affordances into learning technologies to engage young people and support personally relevant learning, but youth may reject these attempts because they do not meet user expectations. Through in-depth case studies, we explore the sociotechnical ecosystems of six teens (ages 15-18) working at a science center that had recently introduced a digital badge system to track and recognize their learning. By analyzing interviews, observations, ecological momentary assessments, and system data, we examined tensions in how badges as connected learning technologies operate in teens' sociotechnical ecosystems. We found that, due to issues of unwanted context collapse and incongruent identity representations, youth only used certain affordances of the system and did so sporadically. Additionally, we noted that some features seemed to prioritize values of adult stakeholders over youth. Using badges as a lens, we reveal critical tensions and offer design recommendations for networked learning technologies.

**CCS CONCEPTS •Human-centered computing~Human computer interaction (HCI)~Empirical studies in HCI•Social and professional topics~User characteristics~Age~Adolescents**

## 1 INTRODUCTION

Young people exist in complex sociotechnical ecosystems that play a variety of roles in their lives, from connecting them with their friends to keeping them apprised of the latest news to tracking their school accomplishments [18,29]. Social media technologies are a significant component of the social and technical worlds in which today's youth interact, experience, and learn. While designers, researchers, and others involved in learning may seek to incorporate social media features and mobile device affordances into various educational technologies and systems, youth may see this as an invasion of what they consider their personal space and culture, with adults imposing rules and restrictions on yet another aspect of their lives [19]. This creates a tension between connecting learning across settings [57] and allowing youth to have spaces in which to develop their own identities, a key part of adolescent development [19,32].

Researchers have found that connecting learning across settings such as school, afterschool, and home can be beneficial in developing interest and identity [4,57]. Designers of learning-related applications and systems have sought to incorporate social sharing aspects into these technologies to better scaffold cross-context connections and support learner identities [5,22]. Scaffolding these learner identities can also help support the development of professional online identities, important to long-term success, as youth move into young adulthood [9,32]. These attempts at incorporating social media aspects are not always successful, however, in part because some youth view them as incongruent with the way they use social media technology and present themselves on social networking sites. Helping to explain this pushback is the present reality that many youth are no longer permitted to have much physical privacy or time alone with their friends due to safety concerns, and thus have moved to digital environments, even creating small clubhouse-like spaces within their larger online ecosystems [18,19,72,73]. Youth may take the view that adding elements associated with school to their limited leisure spaces is yet another attempt by adults to infringe upon what little freedom they have [55,65]. This is particularly salient in contexts such as the COVID-19 pandemic, where many youth were spending more time online and were often largely confined to their homes. As technologies collapse contexts [26,64]–removing social and technological walls that previously existed–and connected learning concepts become more prevalent among users and designers of learning technologies, youth may find themselves without digital places where they feel in control, or constantly moving to new sites and platforms to prevent intrusion [19,26,57]. How can we, as researchers, designers, and teachers, help connect learning experiences and support development of related identities while respecting youth digital spaces?

In the current study, we focus on examining how connected learning technologies might operate in youth's sociotechnical ecosystems, using digital badges as a lens and test case for gaining insight into how learning technologies can be implemented in a way that resonates with rather than threatens youth's existing sociotechnical practices. Badges are web-enabled microcredentials with embedded metadata that provide evidence of learning achievements and can be easily shared across online platforms. With their portability and ease of social media sharing, badges provide an excellent test case for examining how youth incorporate (or do not incorporate) such technologies into their sociotechnical ecosystems. As a relatively new effort in educational contexts, badges are at an interesting inflection point: their promise for giving learners greater control over their learning trajectories is recognized by many, but they have yet to be widely adopted and have therefore not crossed the threshold from potential to real-world value [50]. For badges to ultimately be successful and gain widespread currency, becoming robust connected learning technologies, they must integrate effectively with the existing sociotechnical ecosystems of users, meeting their needs and supporting their desired use cases [76].

To better understand how young people might incorporate technologies such as digital badges into their day-to-day lives and how to design badge systems to better accommodate their needs, we conducted a series of case studies with youth as part of a long-term research project focused on the design and implementation of digital badges in STEM learning. The focal questions of this work are:

1. How do youth collapse and separate their online contexts, particularly around professional/academic vs. personal/social boundaries?
2. What tensions do youth navigate in their sociotechnical daily rounds, particularly surrounding identity and context collapse?
3. How might connected learning technologies such as badges be incorporated into teens' sociotechnical ecosystems in a way that supports their learning and identities, while respecting their boundaries?

Through this work, we explore the complex sociotechnical ecosystems of high school students, identifying key barriers and challenges to youth incorporation of connected learning technologies into their routines. In our analysis, we discovered that badge systems were often geared towards use cases that were not part of youth's daily social media interaction patterns, even though, on the technical front, they were easily shareable on social media and represented the youth's achievements. Our participants carefully curated their social media presentations and partitioned their academic and personal identities, with little overlap. At the same time, the youth were eager to use the affordances (e.g., portability, metadata) and potential currency of digital badges in learning contexts as a way to support their academic and professional progress and display their work to education gatekeepers such as college admissions officers, but were often unsure of how best to do so. For badges and similar systems to support youth's identities as learners and be used to their maximum potential, they must work with youth needs. Thus, we posit that badges might serve as a way to support youth's efforts to present themselves professionally online, connecting contexts that *they* wish to connect. In this way, badges have the potential to play a meaningful role in an expanded view of youth's sociotechnical ecosystems, one that includes online professional identity but does not require connections across all aspects of a youth's ecosystem. The primary contribution of this work is empirical evidence providing insight into the tensions between connected learning and unwanted context collapse in the daily technological patterns of youth. Drawing on this empirical insight, we also contribute design considerations for integrating learning technologies into youth's sociotechnical ecosystems in a way that supports both learner identity and youth agency.

## 2 RELATED WORK

In the recent history of designing learning technologies for young people, there has been a focus on how to facilitate connected learning [4,30,57] but less on the subject of context collapse [26,65], the way in which modern technologies remove the boundaries between settings and groups of people in one's life that typically exist separately offline. The field continues to examine how the complexity of youth's sociotechnical ecosystems affects their engagement with learning and their communities and places, forming their identities as individuals [32]. We build upon this body of work, using digital badges as a case of a potential connected learning technology. We examine how badges intersect with the networked lives of youth and how context collapse plays a role in their adoption.

### 2.1 Identity and learning, online and offline

Youth develop their identities, internal representations of who they are in relation to the world around them, based on a variety of factors, including the media they consume and peer influence [36–38,43,44]. From the Eriksonian perspective, adolescent youth construct their identities and explore how they wish to be perceived by peers, mentors, teachers, family, and other groups [32,37,38,46]. Identity is already layered and nuanced, but social media has added to the complexity, with youth also constructing their identities in a highly networked era [32,43]. Understanding *how* youth use social networking platforms and mobile devices to develop and explore their identities is essential for researchers who wish to develop learning technologies that use the affordances of these sites to engage youth and work with concepts such as *connected learning* (discussed below).

Researchers have found that though there are some general patterns of social media interaction, behavior varies across subtypes of platforms and by individuals [27,28,63,65]. Some youth may have multiple online screennames and roles that they perform depending on the audience. At the same time, many youth feel pressure to present an online identity that mirrors their offline identity, particularly on platforms popular with their peer group, such as Instagram [70]. Even on these apps, however, youth are finding ways to present different aspects of themselves. In recent years, scholars have noted the rise of youth using multiple accounts on the same site or application for different purposes, such as *Finstas* ("fake" Instagram accounts) that allow them to post memes or other content that does not fit their main public-facing persona, and can also be a layer of privacy [33]. Some youth are highly selective about what they post on their public profiles in order to maintain their curated image [63,72,73]. Each platform and application that a youth uses may have a separate purpose.

How youth curate these various social media accounts demonstrates that they are highly aware of the need to present different faces in different contexts. Certain accounts are carefully presented for public (or at least school-wide) consumption, while others are restricted to only close friends [28,91]. Many youth are aware of the consequences of negative social media behavior and use a variety of tactics to ensure that their accounts are hidden from unwanted audiences [18,91]. Parents and other authority figures often encourage this degree of discretion, warning youth of the possible negative consequences of posting certain types of personal content online. Some families go so far as to censor youth profiles and posts or attempt to prevent youth from accessing social media altogether, which can also result in youth creating decoy accounts [18,19,90,91]. In some ways, online spaces have replaced previous teen hangout spots such as malls or outdoor spaces, and youth are often fiercely protective of them [19]. Online discussions and connections have also been found to be both important and beneficial to youth, mostly with offline friends they already know [34,87]. No matter how youth are interacting online, their interactions and choices shape their identities, both private and public.

Learning sciences scholars are keen to understand how to promote learner identities, often science or STEM identities in particular, as strong science identities are associated with long-term academic success [9,11,44,56]. Youth, as learners, develop identities around their relationships with topics, learning environments, and other aspects of the sociocultural learning environment [9,11,44]. For instance, Gee discusses different ways to view identity in education, including affinity groups (interactive communities that build relationships around shared interests and goals), and discusses the different types of identities learners navigate [44]. Affinity groups can span physical and digital spaces, particularly relevant to the current work and how youth navigate identity online [44,54]. Barton et al. have explored how learners develop their science identities, particularly examining the trajectories of middle school girls and low-income urban youth [9,11]. Ito et al. (2015) have also examined how youth engage with civics in a more connected manner than ever before, linking politics to their identities and affinities [58].

In this work, we explore how teens are using social media to both build and present specific identities that may be context-dependent, and how certain learning technologies that collapse contexts, whether they be spatial, social, or temporal [18,26] can clash with this pattern of behavior, creating discordant tension. Additionally, in previous research within our larger badging project, we found that youth were often very concerned about their online profiles and told us that they were often warned about what people could find online, in contrast to the public perception that youth are unaware of potential consequences of their behavior. We extend previous work by examining the tension inherent in designing sociotechnical systems to support youth learning identities across settings while avoiding restricting independent identity development.

## 2.2 Connecting learning with social technologies

Drawing on sociocultural learning ideas and the ecosystems in which people grow and develop [20,88], *connected learning theory* postulates that youth will be more engaged and involved in learning if their experiences are focused on their interests and connected across settings, supported by opportunities and relationships [54,57]. Ito and others stress that out-of-school learning experiences are particularly important for reinforcing concepts from school and that informal learning is key to a well-rounded learning ecology [7,9,57]. Various programs have explored this theoretical positioning, based in sociocultural learning theory, to examine how best to support learners across their various settings [4,5,9,30]. Additionally, scholars have examined the specific role of space and place and how they connect to learning and youth experiences [9,10], as well as how, when, and where youth interact with technologies, media, and sociotechnical systems in their daily routines, referred to as their *daily media rounds* [84,86]. Ecological inquiry and artefact ecologies examine the ecosystems surrounding the design process and how technologies might intermesh and be used in context [15,81]. These concepts have led to investigations into how researchers might leverage the existing technology habits of youth to support learning and learner identities.

The popularity of social media as a form of communication among youth has led to significant research on and implementation of social media style tools and applications that can be used in learning environments to support connected learning [5,12,23,48,57,63]. We refer to this diverse set of platforms, tools, and applications as "connected learning technologies" throughout this work. Some projects use certain affordances and stylings of social media to promote communication and sharing, such as a feed of posts and images [4], and researchers have explored the potential of using social media and networks in the classroom [1,3,48]. Researchers have also designed and implemented technologies like *Science Everywhere* and *Zydeco* for connecting learning contexts [4,21]. *Science Everywhere* explores how youth can connect science across settings and was developed through a co-design process with youth and their families [4,93]. *Zydeco* also connects science across settings, allowing students to transfer their learning from the museum to the classroom using a mobile application [21,60]. These applications restrict user experience through teacher oversight to safeguard student privacy, making them more acceptable to schools in terms of rules and legalities.

In the realm of formal education, learning management systems (LMSs) have also become common and incorporate various affordances associated with social media, such as message boards, private messaging, chats, and collaborative spaces. They also bring grades, analytics, assignments, and course activities to one location [13], but such learning technologies do encounter difficulties in terms of technology access [2]. In contrast to LMSs and other academic applications being encouraged by teachers and administrators as valuable links between school and home, social networking sites are heavily restricted and criticized by school districts and many educators, making the implementation of programs that use such sites difficult [3]. These sites are restricted because of privacy laws and safety considerations, as well as their tendency to distract students from their work [41,66]. This means that while social networking sites are extremely popular with youth, educators and designers have difficulty using them freely for learning applications. In this work, we explore the tensions between social and digital media affordances being employed in educational contexts and the ways in which youth perceive and react to these efforts, extending prior work in this area to better understand youth perspectives [4,5,21,48].

### 2.3 Connecting learning with digital badges

Since 2013, research into digital badges has gained significant traction as various scholars explore the potential of these microcredentials for a variety of applications, from professional development trainings at large companies to helping primary school children learn new skills [14,16,22,71,79]. Badge systems, particularly open badges (badges developed using an open technical standard that is publicly available), are a way of presenting skills in a graphical, digital format with embedded metadata, and their flexibility allows them to be arranged in skill trees and learning pathways [6,39,52]. As such, they have the potential to connect learning across settings and be used as microcredentials, showcasing a broader range of experiences than are typically reflected on a school transcript [40].

While badge systems have a great deal of potential for making learning visible across contexts, badges have run into issues of credibility and utility, as they have only come into public consciousness in recent years [71,77]. Another issue is the broader problem of implementation and uptake of such technologies, which involves some specific challenges in informal learning environments [76]. Introductions of new technologies require an adjustment period and may not be sustainable over the long term, which for a technology that badge evangelists suggest could support life-long learning could be a significant issue [51,53,76]. It is one thing to introduce something like a badge system and work with educators and researchers to implement it, and entirely another to actually get youth and others to incorporate the system into their habits and workflows over time. There is also a concern that badges might perpetuate previously existing inequalities in educational systems, with only those who can access the technology and opportunities able to earn and benefit from them, which is also a concern for many learning technologies [74,77].

In our work, we examine the potential of badges over time and space, as a tool both for tracking learning pathways and for sharing achievements and learning across different settings. This project provides an opportunity to examine the function of a badge system within youth sociotechnical systems, providing unique insight into the functioning of badges on a more holistic level. We build on prior work related to badges, connected learning, and how youth build their identities in the current networked era to better understand youth sociotechnical learning ecologies. The current work provides insights not just for digital badges, but also any connected learning system working with or within social media.

## 3 CASE STUDIES

We conducted six case studies with high school students (age 15-18) to explore how digital badges function in the larger context of young people's media ecologies. Talking to and observing these youth over the course of several months as they engaged with diverse forms of technology across their daily contexts gave us insight into how their habits surrounding technology use interacted with the badge system designed to recognize their learning at a local science museum. Using this type of case study methodology allowed us to examine the youth's sociotechnical lives in a holistic, ecological manner [15,68,81,82].

### 3.1 Project context and setting

The case studies presented in this work are part of a larger multi-year study focused on the design and implementation of digital badging for documenting learning in a youth science interpretation program at a science center in the Pacific Northwest. Researchers have worked closely with the students and staff of the science interpretation program to develop the digital badging system, using participatory design (PD) methods [89]. The badge system serves as both a way of tracking progress through learning pathways within the youth interpretation program and a way to visualize and share achievements outside the program. In addition to the PD work, the research team has also collected data including interviews, surveys, observations, and trainings, all related to the design and implementation of the badge system.

The science interpretation program involves approximately 60-70 high school students (ages 14-18) from a broad metropolitan area with a diverse population. These youth are trained as science interpreters and move through different levels of the program as they gain experience and skills. Youth in the program are initially volunteers and then move to paid positions as they progress through the program curriculum. The program currently has four levels of seniority (referred to in this paper numerically for the purpose of anonymity), each of which is associated with specific learning and performance benchmarks. During their time in the program, the youth learn science and interpersonal skills as they present various exhibits to science center visitors. There are also a variety of social and educational opportunities available to participating youth, such as tutoring, internships, and mentorship from the adult interpreters at the science center, providing a significant non-digital aspect of the connected learning ecosystem.

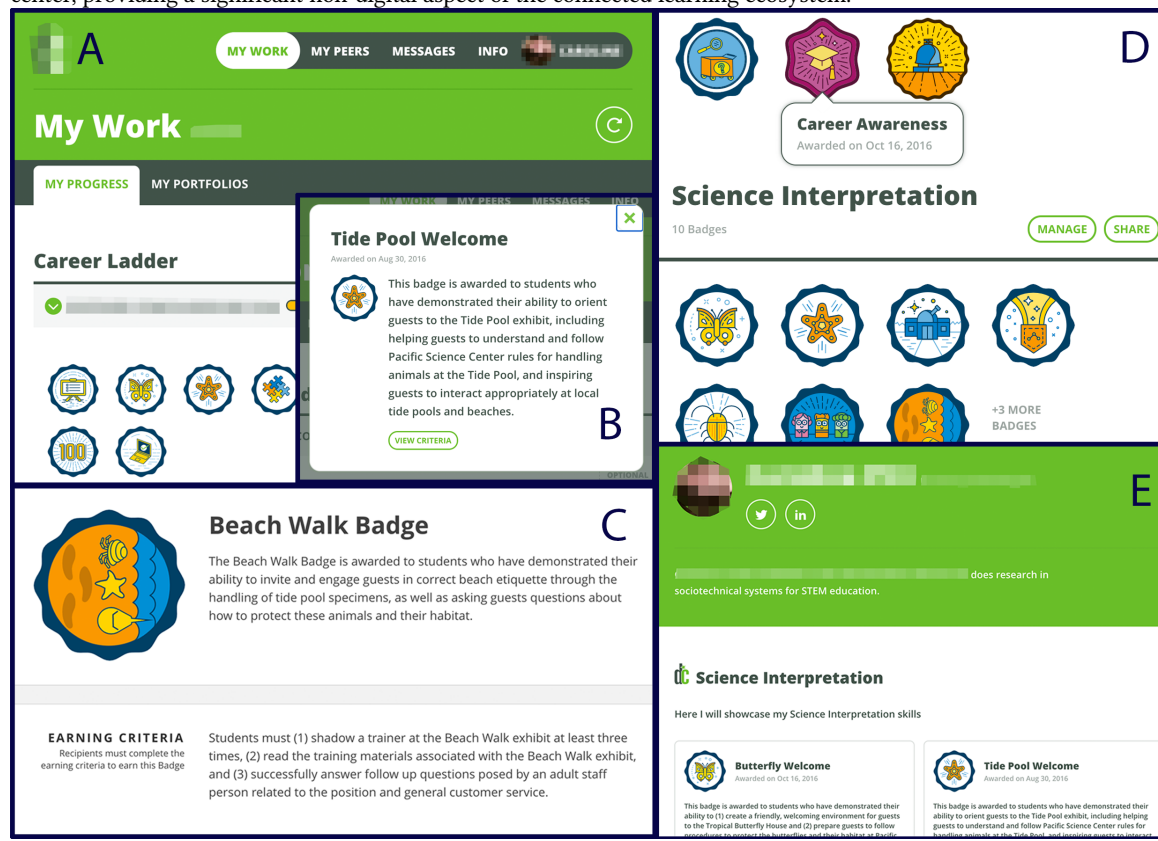

Figure 1. Images of the badge system interface. Top left: A. The "My Work" landing page displaying all badges earned, B. (inset) Badge description pop-up detail for one badge. Bottom left: C. Badge description and criteria for earning a specific badge. Top right: D. Internal view of a portfolio created around showcasing a learner's science interpretation skills. Bottom right: E. External (sharable) view of a portfolio with profile photo, social media links, short biography, and description.

The youth science interpreter program and its four levels of seniority are well-suited to a digital badge system. The clearly defined learning accomplishments associated with each level of the program can be demonstrated and made visible with individual badges that describe exactly what the learner had to do to earn the badge. The badge system that we created with program participants is the result of the design work that has taken place at the science center over the past five years. Working with youth and staff from the science center, the research team conducted a series of design sessions for the initial design of the badge system (focusing on how the badges would be organized, reflect the program curriculum, and so on) and for additional features such as portfolios (collections of badges organized to display to college admissions officers and employers). Youth earn badges by completing certain program benchmarks, or, for certain badge categories, applying for the badges based on accumulated experience and skills. The system allows youth to view their progress through learning pathways (see Figure 1.A), share their badges (for instance on social media or on college applications), create portfolios of badges (Figure 1.D), message their supervisors, view the badge progress of their peers, and create a personal profile with a short biography and photo (Figure 1.E). At the time of these case studies the badge system had been in place and operating in the program for approximately two years. The system is used to track youth progress through the program, and the youth and supervisors are trained in its use.

## 3.2 Participants

All participants were members of the youth science interpretation program at the science center, between 15 and 18 years of age (see Table 1 for details). Participants were recruited with assistance from the program supervisors on the basis of interest in the project and willingness to complete the study procedures. All interested youth received detailed information about the case study procedures before assenting or consenting to the study (those under 18 assented and had parents or guardians sign consent documentation). Ten youth agreed to participate in the case study, but only six were able to finish all procedures, with four either withdrawing for personal reasons or being withdrawn by the research team due to non-responsiveness.

Table 1: Participants, pseudonymized. Level refers to stage of the youth science interpretation program, discussed in section 3.1, ranging from 1 (entry-level) to 4 (most senior and experienced, many do not achieve this level).

| Participant | Grade/Age | Gender | Ethnicity | Level |
|---|---|---|---|---|
| Edith | 12th (17) | Female | Pacific Islander | 3 |
| Alex | 10th (15) | Male | African | 2 |
| Bruce | 12th (18) | Male | White | 4 |
| Eleanor | 12th (17) | Female | Asian | 3 |
| Kelsey | 12th (17) | Female | Asian | 4 |
| Jeremy | 12th (17) | Male | Hispanic | 3 |

## 3.3 Procedure

We selected a case study methodology because it was best suited to provide an in-depth, holistic examination of youth experiences and their interactions with technology in and around the science center [68,69,82,92]. The case studies consisted of an initial interview (one hour), a science center observation (a four-hour shift), a neighborhood interview (approximately two to three hours), ecological momentary assessments (a combination of daily and weekly surveys over multiple weeks), and a final concluding interview (one hour). Thus, approximately ten hours over the course of 4-6 weeks were spent with each participant.

These case studies took place during the 2018-2019 school year and continued into the summer. The majority of the study activities took place at the science center, while neighborhood interviews took place in the greater metropolitan area, in locations chosen by the participants. Three members of the research team (two doctoral students and one master's student) conducted these data collection activities (generally one researcher per activity), under the guidance of the principal investigator. Interviews were recorded and then transcribed. Most science center observations and neighborhood interviews were recorded using a chest- or shoulder-mounted camera [85] and all were documented with extensive field notes and subsequent analytical memos.

*Initial interview.* The initial interview focused on developing an understanding of each participant's motivations and goals, experiences with the science center, science identity, relationship with technology, and knowledge of digital badges [69].

*Science center observation.* To better understand the use of technology and digital badges in particular at the science center, as well as the activities that the participants engaged in on their shifts, researchers observed the youth during a typically four-hour shift at work. Notes were taken using a protocol based in ethnographic methods [45,47].

*Neighborhood interview.* What were intended to be at-home re-enactments [75] were restricted by researcher access to households. While one participant took part in the at-home reenactment, the rest of our participants chose to meet a researcher at a local space that was significant to them (i.e., coffee shop or library). The researchers then conducted a semi-structured interview focused on youth's sociotechnical activities [8,69,86]. Oftentimes participants walked the researchers–physically and verbally–through the space in relation to their daily schedules. Such explorations of online and offline context are extremely important to understanding the ecologies of youth, not only looking at the technologies and systems they use, but the entirety of their social environments [15,81,86].

*Ecological momentary assessments.* The ecological momentary assessments (EMAs) took place over the course of multiple weeks [49,83]. After initial testing with the first three participants, short surveys regarding their recent technology usage were sent once per day (until at least 21 had been completed). These surveys provided snapshots of participants' in-situ technology and science experiences.

*Exit interview.* The final or "exit" interview concluded each case study. This interview reviewed previous interactions, encouraging reflection on the observations and EMAs. Participants were given the opportunity to see and respond to their survey data and reflect on their interactions with their mobile devices as well as the badge system. This interview was also an opportunity for them to provide feedback on the experience of participating in the case study.

## 3.4 Analysis

The analysis of the case study data comprised two main stages: initial writing of detailed memos summarizing the data and then more focused coding and exploration of major themes [17]. The first stage occurred during the summer of 2019 and the second during the fall of 2019, with some additional analysis continuing into 2020.

Shortly after data collection, members of the research team went through the data and engaged in a memo writing process. Three experienced research team members (two PhD candidates and a professor) took different sets of data from the case studies (interviews, observations, and ecological momentary assessments respectively) and reviewed the contents, creating analytical memos and summaries that were then presented and reviewed during project meetings during the summer of 2019 [17,59]. This process highlighted certain key findings and themes from the data, including badge system usage habits and general trends.

In the second stage of data analysis, the first and third authors reviewed the memos and engaged in a collaborative thematic coding process to identify major themes [17,24,69,80] surrounding their perspectives on social media, mobile devices, and educational technology, as well as how digital badges fit into their sociotechnical ecosystems. After deriving the initial themes, the first and third authors went back through the large data corpus of memos, transcripts, and other documentation to revise the coding, coming to consensus on the themes through discussion [62,67,80].

## 4 FINDINGS

Each participant presented a unique and rich perspective on the social media habits and daily rounds of youth [86], particularly youth who are highly engaged in science learning. First, we present brief profiles to provide ecological context about each participant. Then, we discuss the daily rounds of these youth and explore the main tensions and themes that analysis of these case studies surfaced. Overall, while the participants were interested in how digital badges could help them with their learning processes, their social media spheres were largely separated from their learning experiences and they rarely engaged with badges on social media or even outside the science center. Participants identified several reasons for not posting about badges or their academic experiences more broadly. These reasons included unwanted intrusion of school-like experiences on their social spaces (context collapse), authority figures recommending not sharing personal information, and perceived lack of interest from friends and followers.

### 4.1 Participant profiles

#### 4.1.1 Edith – Active social media expert

Edith was a high school senior taking some college classes at the time of the case study. She had also reached the second highest tier of the science interpretation program. Edith felt strong ties to her family and expressed an interest in environmental science, pharmacology, and anthropology. She enjoyed outdoor activities like hiking and working as a camp counselor. Edith also talked about attending church and being on the school's basketball team. She used social media regularly to keep in touch with friends, mostly on Snapchat and Instagram, and checked notifications often. Edith was also involved in the program's media team, which included assisting with the program's Snapchat and Instagram accounts.

#### 4.1.2 Alex – Robotics programmer

Alex was in 10th grade at the time of the case study and was at the second of four levels in the program. He discussed wanting to make his Ethiopian immigrant family proud as part of his core motivation. In terms of social media, Alex enjoyed consuming rather than producing content, looking for reviews of technology online and keeping up on sports. He was in the robotics club and considered himself a programmer, strongly establishing this as part of his identity. Alex was a teaching assistant (TA) for the school robotics class and the only male-presenting individual at the Girls Who Code meeting observed at the Science Center. When not studying at the library or attending robotics club meetings, Alex enjoyed watching soccer matches.

#### 4.1.3 Bruce – Involved athlete

Bruce was in 12th grade, had reached the highest level of the science center program, and had fairly concrete ideas of his interests after high school. He wanted to major in meteorology and Spanish. In addition to his job at the science center, he was also on the baseball team and captained the golf team. When not involved in sports or at the science center, he also attended various church-related activities. His social media and device use mostly focused on keeping in touch with family, peers, and teammates; listening to music; and keeping track of events on various calendar applications. Unlike many of the other participants, he mostly used Facebook Messenger rather than Instagram, Snapchat, or another chat application.

### 4.1.4 Eleanor – Intersectional thinker

Eleanor was a senior, at the second highest level in the program, and taking college classes at the local community college with a focus in social sciences. Her father was a professor and she described her household as scientifically minded. Her interest in social science was broad, "I like thinking about science, how it interacts with society as a whole. So that's where the sociology/anthropology bit comes in." In her interviews she raised a lot of intersectional concerns around race and class, access to opportunities, and academic tracking. She regularly used social media to plan outings with friends and keep up to date on the news.

### 4.1.5 Kelsey – Community-minded traveler

Kelsey, also a senior, did not have to attend full school days any longer due to her senior status. She was at the highest level of the program and enjoyed participating in a large number of community service activities and after-school clubs, including several leadership roles. Kelsey also discussed her love of travel and interest in social sciences. During the case study, she was already discussing her plans for study abroad in university. In terms of social media and technology, she did not have much data included in her phone plan but did maintain many Snapchat streaks and chat conversations when Wi-Fi was available, usually at home.

### 4.1.6 Jeremy – Inspired teacher

Jeremy, a 12[th] grader at the time of the case study, worked both at the science center and at a grocery store as a courtesy clerk. He said that many of the social skills he had learned at the science center were highly transferable. Jeremy mentioned that he had been inspired by various figures in his life to become a counselor or teacher and enjoyed positions at the science center that allowed him to hone his teaching skills. Music featured heavily in Jeremy's routine, with multiple sets of headphones for different environments. He had some social media accounts that he used to talk to a few friends, but mostly he consumed rather than created content, watching YouTube videos for reviews of technology, flipping through Snapchat, and watching soccer.

## 4.2 Daily technology rounds

Each participant followed a pattern of technology use throughout their daily round [86]. Generally, the use of digital technologies fell into three main categories: utilitarian (alarms, calendars, to-do lists, maps, transportation), social (texting, chat apps, social networking sites), and school or work related (email, LMSs, subject or workplace specific applications). Games were less commonly mentioned, but still present. All participants used both a laptop and a phone, with the phone generally used for social activities while the laptop was used primarily for school applications.

To start their day, many, but not all, participants kept their phones next to their beds and used the phone alarm to wake in the morning. Several checked their notifications in the morning, and some listened to music on their devices on the way to school or work. Bruce mentioned that if he forgot his headphones he felt "irritable." Meanwhile, Jeremy had multiple audio devices available, because certain headphones could not charge while in use. Device use on the way to school was very common, with music, bus schedules, social media, and checking to-do lists all being mentioned as part of the routine. As Alex stated about receiving assignments: "I have a to-do list app and I can put it on there. And if there is an event, I usually put it on my calendar."

In the responses to the ecological momentary assessments, boredom was in the top two reasons for checking their device for five out of the six participants, with Alex being the exception. Notably, the most common time to check notifications on devices was during lulls in the day, often in the morning before school or afternoon between activities.

Laptops were taken out for notetaking and classwork, while social networking activities were usually restricted to mobile devices. Participants mentioned that it was very common to use various techniques to get around school rules surrounding device use, however the surveys suggested that much of the device use was before and after school rather than during class time. During her neighborhood interview, Eleanor explained, "Betternet is a VPN because the school blocks stuff so most kids have [it]." Eleanor, Bruce, and Edith all mentioned having VPN apps. Participants mentioned that educators enforced school rules surrounding device use with different degrees of severity. As Jeremy put it, "Yeah [teachers make you put away your phone], but not if you're a senior. They're not really on you as much anymore." While it was a more informal setting, device use at the science center was similar to school, only allowed on breaks or when needed for a task.

Moving towards the end of the day, participants tended to use their devices in transit and sometimes during hangouts with friends, watching videos or listening to music. Several participants stated that they were not allowed to use their phones at family dinners, though Eleanor mentioned that this restriction might be lifted to show an interesting news article or post. There was often a final social media check before sleeping. Jeremy would flip through Snapchat stories before bed, while Kelsey mentioned that most people were not on social media late at night because they were finishing their homework. Do Not Disturb and silent settings were used during class and at night to avoid interruptions (and authority figure detection). Overall, the youth had a fairly strong association of certain tasks at certain times with certain devices, which is also reflected in their social media use.

### 4.3 Social media in tiers and silos

Our participants followed some of the trends documented in other research, such as heavy use of private chat channels with individuals or friend groups and avoiding sharing too much of their online lives with authority figures [18,19,63]. Kelsey noted that sharing too much about her personal or work life was not safe because "you don't wanna reveal too much about yourself on the internet 'cause there's always identity theft and a bunch of other things…" Bruce, who mostly only used Facebook Messenger to chat, had concerns about the safety and security of other applications. While several participants used social networking applications heavily, with Edith and Kelsey both having significant numbers of Snapchat streaks, for example, there were also negative reactions. For instance, Edith, during the course of the case study, mentioned finally breaking her Snapchat streaks because "this one chick she told me that she sets timers just to remember to set Streaks and I was like, 'I don't wanna be you.'"

Each of our participants had very specific audiences in mind for their social media and chat interactions. Eleanor and Kelsey listed specific breakdowns and prioritizations of how close of a friend was contacted on what application, with Eleanor mentioning that texting was reserved for her closest friends, Snapchat was for her broader friend groups, and Instagram was for acquaintances, the widest audience. Kelsey maintained a large number of Snapchat streaks—people that she maintained daily photo exchanges with—(45 at the time of the neighborhood interview) but considered them relatively superficial, using the messaging feature of that application to maintain actual conversations. Kelsey, Eleanor, and Edith also mentioned that they had multiple accounts on Instagram, with one public/real account and at least one other fake/meme account that was usually private and for friends only. Their public accounts usually had fewer posts and were highly curated. The youth did not offer specific reasons for maintaining these divisions, approaching them as a taken-for-granted aspect of online life and easily recounting which groups of friends were on which application.

Youth generally spoke negatively about Facebook ("it's pretty much obsolete," according to Kelsey) and either used it sparingly or not at all, with Bruce the only youth who used it on a regular basis. Edith deactivated her Facebook account and had a Twitter, but only used it for retweeting memes. Meanwhile, Alex used Twitter to keep up on recent technology innovations, and Eleanor used it for world news. Alex and Jeremy were both generally consumers of online content rather than producers of it, mentioning watching soccer matches and YouTube videos but rarely making posts on any social media sites. Each social media site appeared to have a very specific use for these youth, and the audiences and purposes for each account seemed separate and nuanced, even if some of the choices may not have been made explicitly or consciously.

Four of the six youth mentioned using texting or specific chat applications for particularly close friends, while Jeremy preferred to call his girlfriend or speak with her in person. In the ecological momentary assessments, interactions with mobile devices were often rated as meaningful if they involved chatting with friends, rather than scrolling through social media or checking news headlines, weather, or assignments. Several participants mentioned having group chats on specific applications for communicating with their families. Often, these interactions with friends or family involved updating them on current events in the participants' lives or making plans to meet offline.

Across all platforms, the most common use of social media technologies for these youth was private interactions with friends and family with whom they had close relationships offline, whether one-on-one or in small group chats. Very few interactions involved the general public or followers, and such interactions were generally rated as less meaningful in the surveys. Kelsey felt that her time on her phone was least meaningful when she was "just scrolling through social media, or social media that doesn't involve people I know. Just wasting my time on certain stuff," while Eleanor stated that interacting with friends via chat was almost always meaningful and that:

> "I'm hoping that I can get better, especially with my relationship with social media. Especially Instagram, I think I have the worst relationship with, 'cause I go on the most, out of boredom, and it is the least meaningful of any of them, I think."

## 4.4 Identity across contexts and at the science center

These young people had complex lives and constructed identities based on school, hobbies, after-school activities, family, friends, and personal interests [44]. Alex referred to himself as a programmer, Eleanor discussed how her scientifically-minded family was a strong influence on her way of thinking, and Edith felt strong ties to her Polynesian heritage, just to provide a few examples. There was also a substantial online element to participants' identities, as they navigated social media spaces, positioned themselves as for or against certain technologies, and constructed how they wanted to be viewed online, with their various platform profiles. On a meta level, they also contended with the identities placed upon them based on their age and habits [19]. In terms of adult perceptions, Alex both acknowledged them and pushed back, stating that "I see this stereotypical stuff [about teenage social media use], but yeah, I try to keep that out, and manage it in a way that I think it's right." Eleanor felt that "it's very over-simplistic to just be like 'Screenager', 'cause I could very easily be doing something academic [on my phone], organizing for something important in my life." Such statements illustrate youth's awareness of the identities imposed on them by society [59].

In terms of positive influences on identity, most of the participants mentioned their parents, siblings, other relatives, or teachers being good role models and sources of inspiration and motivation. Jeremy, in addition to mentioning teachers, said, "I consider my sister a role model too because she's really bright. [...] She's an architect so I really look up to her." Participants also mentioned specific examples of how these role models helped with their developing identities as emerging adults: Alex, mentioning a robotics teacher who would allow him to take things apart, Eleanor talking about the strength of her grandmother, and Jeremy discussing how his childhood experiences with a counselor made him interested in that profession.

The science center was also often mentioned as a strong and positive source of identity. The members of the science interpretation program were very close, as Edith said, "if we're ever going through something, it's like we can rely on another [member] to help us if we need anything." Alex felt that the program was a great space for like-minded individuals "because it's like there are people who are interested in the same thing, so it would motivate you to pursue that." This statement illustrates how the science program functioned as an affinity space [44] for its youth members, as well as part of the connected learning ecosystem. Program coordinators were also a key element, according to Eleanor:

> "I have a really good relationship with my supervisors. They are mentors. Like I can talk to them about problems. The [supervisor] I was talking to upstairs missed work and he said he felt like he missed his therapy session. It's separate from everything else. School and life and college and this is like a break from everything else."

For many of the students at the science center, the program was their first experience with regular public speaking. Almost all of the participants emphasized this as one of the most important skills learned in the program, transferrable to academic and professional settings. According to Alex, "[the program] changed my social skills and life skills very dramatically, 'cause I can speak to people more." This was a common theme throughout the interviews, with participants coming into their own as teachers, public speakers, and science interpreters through their time at the science center.

In addition to strong bonds, the science interpretation program also encouraged the development of professional and science identities. The program provided opportunities to meet science professionals in various fields, internships, special events and lectures, and affinity-based groups such as the media team and Girls Who Code club. Many of the participants stated that learning at the science center was extremely positive and supportive, as Edith put it: "...let's say you're working on like a science problem with your friend, and instead of them being like "Bruh, how come you don't get it? Like you're stupid?" It's more of like, "Oh my gosh, congrats!"

Although the participants in this study were college-bound and all worked at the science center, they expressed that they did not have much of a professionally oriented online presence and were not as familiar with presenting their professional selves online. When asked if they posted about school or the science center on social media, most said no. Bruce said, "I was posting a little bit about my college stuff on social media, but I didn't really talk about it." Eleanor mentioned that her social media was largely focused on the social: "I don't use a lot of more professional social medias. I don't have a LinkedIn or anything." While the young people involved in this study expressed interest in presenting professional online identities, they had little exposure to this practice. In fact, most of their social media use was locked down or focused on private chats, with very little being intended for a broader public audience.

## 4.5 The case of badges in context

The digital badge system had been part of the science center program for two years at the start of the case studies, and most of the participants were quite familiar with it. Alex thought "the system works well to keep the [program] members well-informed about what they're doing, and it gives us opportunity to share it if we want to." Bruce used the system to check on other members' badges, because "I just think it's cool to see how everyone else is progressing in the program." Despite these stated benefits, Eleanor thought that the badges were somewhat underutilized "cause right now, it's a side thing." Youth noted that the badge system was not checked particularly often due to the fact that supervisors only periodically entered newly-earned badges into the system and because the bulk of the program's workflow remained highly analog. Eleanor suggested that usage could be improved by "having a little more of an incentive for people who want to" use it.

The general consensus seemed to be that the badges were helpful for internal program use (e.g., tracking one's own and others' progress in the program) but less obviously valuable outside the program. Although the youth felt badges were more descriptive than grades, they also felt they were less "serious," which made logging into the system feel less urgent. Comparing the system to school, Edith said:

> "Yeah, with Canvas [LMS], that's something I think about constantly, just because I want to check my work, my grades. Even though this is something that shows you your progress, it's not something that I think, 'Wow, oh my gosh, I need to check this'"

When asked if they would share badges with their friends on social media, each participant had a slightly different perspective. Some contemplated the idea and mentioned that they might in certain circumstances, but a general consensus emerged that there was a strong line between academics and social media. As Edith put it, "Even if my friends were to view [my badges on social media], I don't see them going in depth. I feel them scrolling and they'll be, 'Bro, what the fuck?'" Here, Edith articulates her thoughts about curating online identity, positioning badges and the science center as separate from what she shares with friends. In this case, Edith was referring to school friends who are not particularly familiar with, nor, in her opinion, interested in her science center achievements. The reaction she anticipates from her friends reflects the incongruency of bringing a learning-oriented identity into a distinctly peer-based, social space.

The badge website allowed for the creation of badge portfolios (see Figure 1.D and E) that could be shared externally with stakeholders such as college admissions officers and employers. Jeremy told us that he made use of this feature and shared his badges when he applied for a college scholarship, stating that it was a good way to show what he had done in the program. Other participants acknowledged the potential value of sharing portfolios with external stakeholders, but they had not created any portfolios at the time of the study. Some youth noted that since the potential of badges was not yet a reality or even a recognized added value for these stakeholders, they did not see the need to create a portfolio and share it externally. Even for those youth who were open to sharing portfolios, they expressed uncertainty about how and where to share badges. They recognized that badges could be included as supplemental material in a college application, for instance but they were not sure how the badges would be interpreted by reviewers. Edith recalled that, "When I applied to Howard [University] they were like, oh do you have any social media links that you would like to share to give us a better understanding of who you are?" but she was not sure if this "counted" as a social media link. Her confusion seemed to stem from the difficulty of defining boundaries between professional/academic identity practices and personal social media ones. Given that these youth did not seem to share much of their academic identities online, were unsure of how best to use badges (or other connected learning technologies) to support their external academic and professional identities, and had limited experience with professional and academic online presentation, there is an opportunity for designers, researchers, and educators to create more supportive systems.

## 5 DISCUSSION AND IMPLICATIONS

With the increasing prevalence of digital tools for tracking all aspects of learning and credentialing, learners are technically able to present their skills and achievements online to a variety of audiences. However, they must have personally relevant reasons, motivations, and an appropriate space in which to do so. In this work, we explore the multidimensional sociotechnical experiences of teens in a science interpretation program, specifically looking at how they might integrate a learning technology such as digital badges into their pre-existing media ecologies. We discovered that while youth were interested in tracking their academic achievements online, they did not see much point in sharing their badges (and the experiences they represented) with their friends, families, or online followers. They felt that badges were best used to track and visualize their academic achievements [51,53], rather than presenting such credentials to these more social audiences. Connecting their out-of-school and in-school learning experiences to each other felt natural to them, such as sharing the badges on scholarship and college applications as Jeremy did [30,57], but extending that connection to the spheres of social media and friends felt incongruous. Badges didn't fit their current online profiles, (see sections 4.1-4.3) which showcased carefully curated social identities for an audience of friends and acquaintances [18,32,33,65]. The youth in this study reflected a clear case of pushback against context collapse, with Edith explicitly stating that posting badges to her social feed would be met with a negative reaction from her friends (see 4.5) [19,55,65].

The phenomenon of context collapse has become more prevalent as social media has become a major source of connection and networking, both professionally and socially. Davis and Jurgenson (2014) explore *context collusions*, purposeful collapsing of contexts, versus *context collisions*, which are unwanted and unintentional instances of context collapse [26]. Our case study findings show how youth create their own context collusions by forming categories and tiers for their social media ecologies and maintaining clear, implicit boundaries for certain aspects of their identity representations [19,63]. To this end, designers should seek ways to facilitate youth's context collusions, helping youth *control* the ways they connect aspects of their online identities. Our work suggests that with such support, youth would be more open to using connected learning technologies such as badges in the future. The youth in the current study expressed interest in using their badges for academic and professional applications, yet they were unsure of how and where to use them, indicating an interest in *context collusion* but a need for support to fully enact it.

We see great potential in striving to overcome the challenge of unwanted context collapse between youth's social and educational worlds online. In addition to their social online identities—which are typically well-developed—youth are continuing to develop their identities as learners and figuring out how to present themselves professionally (and as emerging adults) as they grow up in this networked era [19,32,65]. While young people often receive training and parental guidance on what *not* to do or share online [25,90] and how to present themselves professionally *on paper* and *in person* (via school, family, internships, and so on), tools for teaching them how to develop their identities as professional adults in *online* spaces are not nearly as common. Digital badges and badge systems represent an opportunity to scaffold this professional identity development by providing a structured, yet customizable, platform for youth to make their learning and accomplishments visible [22,52,53,71,77]. The current work suggests what is clearly needed going forward is adequate support and guidance (both technical and social) in helping youth decide where, when, and with whom to share their learning accomplishments.

Connected learning strives to encourage deeper, interest-driven learning by promoting connections among youth's experiences in various settings [30,57]. While context collapse via sociotechnical systems can facilitate this, it can also create some value tensions [42] and contradictions, as we discovered when asking participants about how badges interacted with their social media. If we see this tension point as an opportunity to refine the ways we implement sociotechnical systems for connecting learning across settings, we can create tools that allow youth to practice, hone, and develop professional identities and broader identity relations. Based on this opportunity and our case study findings, we propose the following suggestions for designers and researchers developing connected learning technologies, particularly as a tool for youth academic and professional identity development:

1. *When involving youth, bring in the entirety of their experience.* The current study gave us insight into the many different contexts shaping youth's media ecosystems, as well as the highly differentiated ways that individual youth make use of digital tools and platforms. It is important to make sure that their ecologies across time, space, and place are considered in the design of connected learning technologies [81]. The youth experience is best explained by the youth themselves, as discussed in previous work [35,78,89], thus researchers and designers should strive to involve and work with youth beyond just a few design sessions. Specifically, it is important to consider dimensions of teens' lived experiences that extend beyond the design project at hand, drawing on concepts like ecological inquiry [81].

2. *Design for context collusion, not collision.* Youth may avoid sharing badges on social media, as discussed and explored in section 4.5, but they expressed interest in including them in academic and professional areas, such as internship and college applications. The participants wanted to transfer and translate their learning across settings, but in a controlled way, for specific audiences. Designers should consider where and how to promote context collusion, without intruding on spaces youth wish to keep separate [26,54,65]. Participants curated their public posts and were concerned about privacy, taking care to keep identifying information off of certain accounts. Building on previous work regarding how youth form different identities in different contexts [19,32,65,72,73], we suggest that designers create tools that safeguard youth privacy but allow the creation of public-facing pages, profiles, or other displays of the youth's achievements. Adding features like different display and embed formats and the ability to hide or display certain information easily are also features desired by youth participants, as there were many different places they might share the information. The key is to make sure youth are able to engage with professional identity in a controlled manner that avoids context collision (and, of course, maintains privacy and safety).

3. *Scaffold development of professional identity while allowing for personal customization.* As discussed in section 4.4, youth wanted to develop their professional identities but had varying levels of experience with professional online networking tools. They had heard of sites like LinkedIn but had not used them. Youth have considerable expertise in social media, but not for such contexts, so consider how connected learning technologies can translate that to a more academic or professional presentation. For instance, for something like a badge system, we suggest making it easy to include the profile, badges, or other information on a resume or college application in a format that is appropriately professional, as youth can be concerned with meeting professional norms [31]. By providing scaffolding such as suggestions and outlines in an informal learning space, designers and educators can assist youth in developing profiles appropriate for sharing with academic and professional contexts.

These recommendations provide insights into how networked technologies and tools might be used for connecting learning and scaffolding youth's online academic and professional identity development while allowing youth to maintain their own online spaces. This is particularly important for adolescents, who have very few spaces–online and off–to themselves, but who are also beginning to join professional spaces as they move towards the public, professional life of adults [19]. Our findings indicate that youth are highly resistant to context collisions but may be in favor of context collusions [26], providing an opportunity for sharing achievements and encouraging connected learning across some, but not all, settings. These insights suggest that tools like digital badges would be well-suited to helping youth build professional identity presentations using learning technologies as a scaffolding mechanism, supporting youth as they move through the complexity of adolescence and the many identities and spaces they navigate.

## 6 LIMITATIONS AND FUTURE WORK

This work is limited in explanatory scope due to the case study methodology, particularly given a small sample from a science center program in a major metropolitan area. However, such design allows for deep examination of the particular cases and an ecological perspective [81], focusing on the theoretical and practical tensions at play. Despite the small sample size, these holistic explorations provided a more complete view of the sociotechnical ecosystems of the youth involved, especially in terms of their learning experiences. The case study data is also part of a larger ongoing study, and the design draws on a significant corpus of data and experience across participants and researchers. With respect to generalizability, we might expect similar results in structured youth programs in libraries, museums, recreation centers, and other comparable spaces. Like the science center, these settings are more aligned with learning than the peer cultures found on sites like Instagram and Snapchat, and studies consistently find youth endeavoring to keep these contexts separate.

In the future, we hope to examine youth technology ecosystems more broadly in terms of connected learning and contextual identities, across more settings and demographics. We also plan to further investigate designing to support youth in their exploration of professional identity, figuring out how and to what degree to present themselves publicly as professionals online. We hope to gather more insight on how to mediate the tensions between context collapse and connected learning, creating sociotechnical systems that provide balance for the networked lives of teens.

## 7 CONCLUSION

Connected learning technologies and context collapse are inextricably intertwined, creating difficult tensions for those who wish to allow youth places to develop their own personal and social identities while assisting them in developing their academic and professional ones. In this study, we examine the sociotechnical ecosystems of youth science interpreters, exploring how their use of social media technologies contributes to their identities and how they resist context collapse. Although this resistance creates challenges for connected learning technologies such as digital badges, we suggest how designers can approach the development and implementation of these technologies in a way that respects the needs of young people, including the need to maintain boundaries in and across online spaces, and helps to scaffold their developing professional identities online.

## ACKNOWLEDGEMENTS

The authors would like to thank the science center and the case study participants, as well as the research team.

The work is supported by the National Science Foundation under Grant No.: DRL-1452672 (https://www.nsf.gov/awardsearch/showAward?AWD_ID=1452672).

## REFERENCES

[1]  Paige Abe and Nickolas A. Jordan. 2013. Integrating Social Media Into the Classroom Curriculum. *About Campus* 18, 1: 16–20. https://doi.org/10.1002/abc.21107

[2]  June Ahn, Austin Beck, John Rice, and Michelle Foster. 2016. Exploring Issues of Implementation, Equity, and Student Achievement With Educational Software in the DC Public Schools. *AERA Open* 2, 4: 2332858416667726. https://doi.org/10.1177/2332858416667726

[3]  June Ahn, Lauren K. Bivona, and Jeffrey DiScala. 2011. Social media access in K-12 schools: Intractable policy controversies in an evolving world. *Proceedings of the American Society for Information Science and Technology* 48, 1: 1–10. https://doi.org/10.1002/meet.2011.14504801044

[4]  June Ahn, Tamara Clegg, Jason Yip, Elizabeth Bonsignore, Daniel Pauw, Lautaro Cabrera, Kenna Hernly, Caroline Pitt, Kelly Mills, Arturo Salazar, Diana Griffing, Jeff Rick, and Rachael Marr. 2018. Science Everywhere: Designing Public, Tangible Displays to Connect Youth Learning Across Settings. In *Proceedings of the 2018 CHI Conference on Human Factors in Computing Systems* (CHI '18), 278:1–278:12. https://doi.org/10.1145/3173574.3173852

[5]  June Ahn, Tamara Clegg, Jason Yip, Elizabeth Bonsignore, Daniel Pauw, Michael Gubbels, Charley Lewittes, and Emily Rhodes. 2014. Seeing the unseen learner: designing and using social media to recognize children's science dispositions in action. *Learning, Media and Technology* 0, 0: 1–31. https://doi.org/10.1080/17439884.2014.964254

[6]  June Ahn, Anthony Pellicone, and Brian S. Butler. 2014. Open badges for education: what are the implications at the intersection of open systems and badging? *Research in Learning Technology* 22, 0. https://doi.org/10.3402/rlt.v22.23563

[7]  Brigid Barron. 2006. Interest and Self-Sustained Learning as Catalysts of Development: A Learning Ecology Perspective. *Human Development* 49, 4: 193–224. https://doi.org/10.1159/000094368

[8]  Brigid Barron, Amber Levinson, Caitlin K Martin, Veronique Mertl, Daniel Stringer, Kimberly Austin, Nichole Pinkard, Kimberly Richards, and

Kimberly Gomez. 2010. Supporting young new media producers across learning spaces: a longitudinal study of the digital youth network. 2: 8.

[9] Angela Calabrese Barton, H. Kang, E. Tan, T. B. O'Neill, J. Bautista-Guerra, and C. Brecklin. 2013. Crafting a Future in Science: Tracing Middle School Girls' Identity Work Over Time and Space. *American Educational Research Journal* 50, 1: 37–75. https://doi.org/10.3102/0002831212458142

[10] Angela Calabrese Barton and Edna Tan. 2009. Funds of knowledge and discourses and hybrid space. *Journal of Research in Science Teaching* 46, 1: 50–73. https://doi.org/10.1002/tea.20269

[11] Angela Calabrese Barton and Edna Tan. 2010. We Be Burnin'!: Agency, Identity, and Science Learning. *Journal of the Learning Sciences* 19, 2: 187–229. https://doi.org/10.1080/10508400903530044

[12] Adam Bell, Katie Headrick Taylor, Erin Riesland, and Maria Hays. 2019. Learning to see the familiar: Technological assemblages in a higher education (non) classroom setting. *British journal of educational technology* 50, 4: 1573–1588.

[13] Paulo Blikstein and Marcelo Worsley. 2016. Multimodal Learning Analytics and Education Data Mining: using computational technologies to measure complex learning tasks. *Journal of Learning Analytics* 3, 2: 220–238.

[14] Goldie Blumenstyk. 2018. With Employers in the Mix, Can Badges Become More Than a Fad? *The Chronicle of Higher Education.* Retrieved September 11, 2018 from https://www.chronicle.com/article/With-Employers-in-the-Mix-Can/244322

[15] Susanne Bødker and Clemens Nylandsted Klokmose. 2012. Dynamics in artifact ecologies. In *Proceedings of the 7th Nordic Conference on Human-Computer Interaction: Making Sense Through Design* (NordiCHI '12), 448–457. https://doi.org/10.1145/2399016.2399085

[16] Ivica Boticki, Jelena Baksa, Peter Seow, and Chee-Kit Looi. 2015. Usage of a mobile social learning platform with virtual badges in a primary school. *Computers & Education* 86: 120–136. https://doi.org/10.1016/j.compedu.2015.02.015

[17] Richard E. Boyatzis. 1998. *Transforming qualitative information: Thematic analysis and code development.* SAGE.

[18] danah boyd. 2007. *Why Youth (Heart) Social Network Sites: The Role of Networked Publics in Teenage Social Life.* Social Science Research Network, Rochester, NY. Retrieved April 30, 2019 from https://papers.ssrn.com/abstract=1518924

[19] danah boyd. 2014. *It's complicated: The social lives of networked teens.* Yale University Press.

[20] Urie Bronfenbrenner. 1977. Toward an experimental ecology of human development. *American psychologist* 32, 7: 513.

[21] Clara Cahill, Alex Kuhn, Shannon Schmoll, Alex Pompe, and Chris Quintana. 2010. Zydeco: Using Mobile and Web Technologies to Support Seamless Inquiry Between Museum and School Contexts. In *Proceedings of the 9th International Conference on Interaction Design and Children* (IDC '10), 174–177. https://doi.org/10.1145/1810543.1810564

[22] Carla Casilli and Daniel Hickey. 2016. Transcending conventional credentialing and assessment paradigms with information-rich digital badges. *The Information Society* 32, 2: 117–129. https://doi.org/10.1080/01972243.2016.1130500

[23] Tamara Clegg, Jason C. Yip, June Ahn, Elizabeth Bonsignore, Michael Gubbels, Becky Lewittes, and Emily Rhodes. 2013. When face-to-face fails: Opportunities for social media to foster collaborative learning. In *Tenth international conference on computer supported collaborative learning.*

[24] Juliet Corbin and Anselm Strauss. 2015. *Basics of Qualitative Research: Techniques and Procedures for Developing Grounded Theory.* SAGE.

[25] Alexei Czeskis, Ivayla Dermendjieva, Hussein Yapit, Alan Borning, Batya Friedman, Brian Gill, and Tadayoshi Kohno. 2010. Parenting from the Pocket: Value Tensions and Technical Directions for Secure and Private Parent-teen Mobile Safety. In *Proceedings of the Sixth Symposium on Usable Privacy and Security* (SOUPS '10), 15:1–15:15. https://doi.org/10.1145/1837110.1837130

[26] Jenny L. Davis and Nathan Jurgenson. 2014. Context collapse: theorizing context collusions and collisions. *Information, Communication & Society* 17, 4: 476–485. https://doi.org/10.1080/1369118X.2014.888458

[27] Katie Davis. 2012. Tensions of identity in a networked era: Young people's perspectives on the risks and rewards of online self-expression. *New Media & Society* 14, 4: 634–651.

[28] Katie Davis. 2012. Friendship 2.0: Adolescents' experiences of belonging and self-disclosure online. *Journal of adolescence* 35, 6: 1527–1536.

[29] Katie Davis, Anja Dinhopl, and Alexis Hiniker. 2019. "Everything's the Phone": Understanding the Phone's Supercharged Role in Parent-Teen Relationships. In *Proceedings of the 2019 CHI Conference on Human Factors in Computing Systems - CHI '19,* 1–14. https://doi.org/10.1145/3290605.3300457

[30] Katie Davis and Sean Fullerton. 2016. Connected learning in and after school: Exploring technology's role in the learning experiences of diverse high school students. *The Information Society* 32, 2: 98–116. https://doi.org/10.1080/01972243.2016.1130498

[31] Katie Davis, Caroline Pitt, Adam Bell, and Ada Kim. 2018. Using digital badges to promote student agency and identity in science learning. In *Proceedings of the Connected Learning Summit (CLS '18).*

[32] Katie Davis and Emily Weinstein. 2017. Identity development in the digital age: An Eriksonian perspective. In *Identity, sexuality, and relationships among emerging adults in the digital age.* IGI Global, 1–17.

[33] Sofia Dewar, Schinria Islam, Elizabeth Resor, and Niloufar Salehi. 2019. Finsta: Creating "Fake" Spaces for Authentic Performance. In *Extended Abstracts of the 2019 CHI Conference on Human Factors in Computing Systems - CHI EA '19,* 1–6. https://doi.org/10.1145/3290607.3313033

[34] Michal Dolev-Cohen and Azy Barak. 2013. Adolescents' use of Instant Messaging as a means of emotional relief. *Computers in Human Behavior* 29: 58–63. https://doi.org/10.1016/j.chb.2012.07.016

[35] Allison Druin. 2002. The role of children in the design of new technology. *Behaviour & Information Technology* 21, 1: 1–25. https://doi.org/10.1080/01449290110108659

[36] Lia Emanuel and Danaë Stanton Fraser. 2014. Exploring Physical and Digital Identity with a Teenage Cohort. In *Proceedings of the 2014 Conference on Interaction Design and Children* (IDC '14), 67–76. https://doi.org/10.1145/2593968.2593984

[37] Erik H. Erikson. 1968. *Identity: Youth and crisis.* WW Norton & Company.

[38] Erik H. Erikson. 1980. *Identity and the life cycle*. Norton, New York, NY.

[39] Erin Fields. 2015. Making Visible New Learning: Professional Development with Open Digital Badge Pathways. *Partnership: the Canadian Journal of Library and Information Practice and Research* 10, 1: 1–10.

[40] Barry Fishman, Stephanie Teasley, and Steven Cederquist. 2018. *Micro-Credentials as Evidence for College Readiness: Report of an NSF Workshop*. Retrieved from https://deepblue.lib.umich.edu/handle/2027.42/143851

[41] Abraham E. Flanigan and Wayne A. Babchuk. 2015. Social media as academic quicksand: A phenomenological study of student experiences in and out of the classroom. *Learning and Individual Differences* 44: 40–45. https://doi.org/10.1016/j.lindif.2015.11.003

[42] Batya Friedman and David G. Hendry. 2019. *Value Sensitive Design: Shaping Technology with Moral Imagination*. MIT Press.

[43] Howard Gardner and Katie Davis. 2013. *The app generation: How today's youth navigate identity, intimacy, and imagination in a digital world*. Yale University Press.

[44] James Paul Gee. 2000. Identity as an Analytic Lens for Research in Education. *Review of Research in Education* 25: 99–125. https://doi.org/10.2307/1167322

[45] Clifford Geertz. 1973. Thick Description: Toward an Interpretive Theory of Culture. In *The Interpretation of Cultures*. Basic books.

[46] Erving Goffman. 1956. *The presentation of self in everyday life*. University of Edinburgh, Edinburgh.

[47] Erving Goffman. 1989. On fieldwork. *Journal of contemporary ethnography* 18, 2: 123–132.

[48] Christine Greenhow and Cathy Lewin. 2016. Social media and education: reconceptualizing the boundaries of formal and informal learning. *Learning, Media and Technology* 41, 1: 6–30. https://doi.org/10.1080/17439884.2015.1064954

[49] Kristin E. Heron and Joshua M. Smyth. 2010. Ecological momentary interventions: Incorporating mobile technology into psychosocial and health behaviour treatments. *British Journal of Health Psychology* 15, 1: 1–39. https://doi.org/10.1348/135910709X466063

[50] Daniel Hickey. 2017. How Open E-Credentials Will Transform Higher Education. *The Chronicle of Higher Education*. Retrieved September 11, 2018 from https://www.chronicle.com/article/How-Open-E-Credentials-Will/239709

[51] Daniel T. Hickey and Grant T. Chartrand. 2019. Recognizing competencies vs. completion vs. participation: Ideal roles for web-enabled digital badges. *Education and Information Technologies*. https://doi.org/10.1007/s10639-019-10000-w

[52] Daniel T. Hickey and Katerina Schenke. 2019. Open Digital Badges and Reward Structures. . Cambridge University Press. Retrieved November 15, 2019 from https://scholarworks.iu.edu/dspace/handle/2022/22941

[53] Daniel T Hickey and James E. Willis. 2017. *Where Open Badges Appear to Work Better: Findings from the Design Principles Documentation Project*. Indiana University, Center for Research on Learning and Technology. Retrieved from https://www.researchgate.net/profile/Daniel_Hickey2/publication/312740974_Design_Principles_for_Digital_Badge_Systems/links/5a818eb10f7e9be137ca69af/Design-Principles-for-Digital-Badge-Systems.pdf

[54] M. Ito, R. Arum, D. Conley, K. Gutiérrez, B. Kirshner, S. Livingstone, V. Michalchik, W. R. Penuel, K. Peppler, and N. Pinkard. 2020. *The Connected Learning Research Network: Reflections on a decade of engaged scholarship*. Connected Learning Alliance. Retrieved from https://clalliance.org/wp-content/uploads/2020/02/CLRN_Report.pdf

[55] Mizuko Ito, Sonja Baumer, Matteo Bittanti, danah boyd, Rachel Cody, Becky Herr Stephenson, Heather A. Horst, Patricia G. Lange, Dilan Mahendran, Katynka Z. Martínez, C. J. Pascoe, Dan Perkel, Laura Robinson, Christo Sims, and Lisa Tripp. 2019. *Hanging Out, Messing Around, and Geeking Out: Kids Living and Learning with New Media*. MIT Press.

[56] Mizuko Ito, Sonja Baumer, Matteo Bittanti, Rachel Cody, Becky Herr Stephenson, Heather A. Horst, Patricia G. Lange, Dilan Mahendran, Katynka Z. Martínez, C. J. Pascoe, Perkel, Dan, Robinson, Laura, Sims, Christo, and Tripp, Lisa. 2009. *Hanging out, messing around, and geeking out: Kids living and learning with new media*. MIT press.

[57] Mizuko Ito, Kris Gutiérrez, Sonia Livingstone, Bill Penuel, Jean Rhodes, Katie Salen, Juliet Schor, Julian Sefton-Green, and S. Craig Watkins. 2013. *Connected learning: An agenda for research and design*. Digital Media and Learning Research Hub, Irvine, CA, USA. Retrieved October 24, 2015 from http://dmlhub.net/

[58] Mizuko Ito, Elisabeth Soep, Neta Kligler-Vilenchik, Sangita Shresthova, Liana Gamber-Thompson, and Arely Zimmerman. 2015. Learning connected civics: Narratives, practices, infrastructures. *Curriculum Inquiry* 45, 1: 10–29.

[59] Brigitte Jordan and Austin Henderson. 1995. Interaction Analysis: Foundations and Practice. *Journal of the Learning Sciences* 4, 1: 39–103. https://doi.org/10.1207/s15327809jls0401_2

[60] Alex Kuhn, Brenna McNally, Shannon Schmoll, Clara Cahill, Wan-Tzu Lo, Chris Quintana, and Ibrahim Delen. 2012. How Students Find, Evaluate and Utilize Peer-collected Annotated Multimedia Data in Science Inquiry with Zydeco. In *Proceedings of the SIGCHI Conference on Human Factors in Computing Systems* (CHI '12), 3061–3070. https://doi.org/10.1145/2207676.2208719

[61] Simone Lanette, Phoebe K. Chua, Gillian Hayes, and Melissa Mazmanian. 2018. How Much is "Too Much"? The Role of a Smartphone Addiction Narrative in Individuals' Experience of Use. *Proceedings of the ACM on Human-Computer Interaction* 2, CSCW: 101:1–101:22. https://doi.org/10.1145/3274370

[62] Heidi M Levitt, Michael Bamberg, John W Creswell, David M Frost, Ruthellen Josselson, and Carola Suárez-Orozco. 2018. Journal article reporting standards for qualitative primary, qualitative meta-analytic, and mixed methods research in psychology: The APA Publications and Communications Board task force report. *American Psychologist* 73, 1: 26.

[63] Adriana M. Manago. 2014. Identity development in the digital age: The case of social networking sites. *The Oxford handbook of identity development*: 508–524.

[64] Alice E. Marwick and danah boyd. 2011. I tweet honestly, I tweet passionately: Twitter users, context collapse, and the imagined audience. *New*

*Media & Society* 13, 1: 114–133. https://doi.org/10.1177/1461444810365313

[65] Alice E Marwick and danah boyd. 2014. Networked privacy: How teenagers negotiate context in social media ,
Networked privacy: How teenagers negotiate context in social media. *New Media & Society* 16, 7: 1051–1067. https://doi.org/10.1177/1461444814543995

[66] Bernard McCoy. 2016. Digital Distractions in the Classroom Phase II: Student Classroom Use of Digital Devices for Non-Class Related Purposes. *Faculty Publications, College of Journalism & Mass Communications.* Retrieved from https://digitalcommons.unl.edu/journalismfacpub/90

[67] Nora McDonald, Sarita Schoenebeck, and Andrea Forte. 2019. Reliability and Inter-rater Reliability in Qualitative Research: Norms and Guidelines for CSCW and HCI Practice. *Proceedings of the ACM on Human-Computer Interaction* 3, CSCW: 1–23.

[68] Sharan B. Merriam. 1985. The case study in educational research: A review of selected literature. *The Journal of Educational Thought (JET)/Revue de la Pensée Educative*: 204–217.

[69] Sharan B. Merriam. 2009. *Qualitative Research: A Guide to Design and Implementation.* John Wiley & Sons.

[70] Sarah Nicola Metcalfe and Anna Llewellyn. 2020. "It's Just the Thing You Do": Physical and Digital Fields, and the Flow of Capital for Young People's Gendered Identity Negotiation. *Journal of Adolescent Research* 35, 1: 84–110. https://doi.org/10.1177/0743558419883359

[71] Lin Y. Muilenburg and Zane L. Berge. 2016. *Digital Badges in Education: Trends, Issues, and Cases.* Routledge.

[72] Jacqueline Nesi, Sophia Choukas-Bradley, and Mitchell J. Prinstein. 2018. Transformation of adolescent peer relations in the social media context: Part 1—a theoretical framework and application to dyadic peer relationships. *Clinical child and family psychology review* 21, 3: 267–294.

[73] Jacqueline Nesi, Sophia Choukas-Bradley, and Mitchell J. Prinstein. 2018. Transformation of Adolescent Peer Relations in the Social Media Context: Part 2—Application to Peer Group Processes and Future Directions for Research. *Clinical Child and Family Psychology Review* 21, 3: 295–319. https://doi.org/10.1007/s10567-018-0262-9

[74] Michael Olneck. 2015. Whom Will Digital Badges Empower? Sociological Perspectives on Digital Badges. 5–11. Retrieved October 19, 2016 from http://ceur-ws.org/Vol-1358/paper1.pdf

[75] Sarah Pink and Kerstin Leder Mackley. 2014. Re-enactment methodologies for everyday life research: art therapy insights for video ethnography. *Visual Studies* 29, 2: 146–154.

[76] Caroline Pitt, Adam Bell, Edgar Onofre, and Katie Davis. 2019. A Badge, Not a Barrier: Designing for–and Throughout–Digital Badge Implementation. In *CHI Conference on Human Factors in Computing Systems Proceedings,* 14. https://doi.org/10.1145/3290605.3300920

[77] Caroline Pitt, Adam Bell, Rose Strickman, and Katie Davis. 2018. Supporting learners' STEM-oriented career pathways with digital badges. *Information and Learning Science.* https://doi.org/10.1108/ILS-06-2018-0050

[78] Caroline Pitt and Katie Davis. 2017. Designing Together?: Group Dynamics in Participatory Digital Badge Design with Teens. In *Proceedings of the 2017 Conference on Interaction Design and Children* (IDC '17), 322–327. https://doi.org/10.1145/3078072.3079716

[79] Michelle M. Riconscente, Amy Kamarainen, and Margaret Honey. 2013. *STEM badges: Current terrain and the road ahead.* Retrieved October 19, 2016 from https://badgesnysci.files.wordpress.com/2013/08/nsf_stembadges_final_report.pdf

[80] Peter Smagorinsky. 2008. The Method Section as Conceptual Epicenter in Constructing Social Science Research Reports. *Written Communication* 25, 3: 389–411. https://doi.org/10.1177/0741088308317815

[81] Rachel C. Smith, Ole S. Iversen, Thomas Hjermitslev, and Aviaja B. Lynggaard. 2013. Towards an ecological inquiry in child-computer interaction. In *Proceedings of the 12th International Conference on Interaction Design and Children* (IDC '13), 183–192. https://doi.org/10.1145/2485760.2485780

[82] Robert E. Stake. 1995. *The Art of Case Study Research.* SAGE.

[83] Arthur A. Stone, Saul Shiffman, Audie A. Atienza, Linda Nebeling, A. Stone, S. Shiffman, A. A. Atienza, and L. Nebeling. 2007. Historical roots and rationale of ecological momentary assessment (EMA). In *The science of real-time data capture: Self-reports in health research.* Oxford University Press, 3–10.

[84] Katie Headrick Taylor and Rogers Hall. 2013. Counter-mapping the neighborhood on bicycles: Mobilizing youth to reimagine the city. *Technology, Knowledge and Learning* 18, 1–2: 65–93.

[85] Katie Headrick Taylor and Deborah Silvis. 2017. Mobile City Science: Technology-Supported Collaborative Learning at Community Scale. Retrieved January 23, 2020 from https://repository.isls.org//handle/1/256

[86] Katie Headrick Taylor, Lori Takeuchi, and Reed Stevens. 2018. Mapping the daily media round: novel methods for understanding families' mobile technology use. *Learning, Media and Technology* 43, 1: 70–84. https://doi.org/10.1080/17439884.2017.1391286

[87] Patti M. Valkenburg and Jochen Peter. 2009. Social consequences of the Internet for adolescents: A decade of research. *Current directions in psychological science* 18, 1: 1–5.

[88] Lev S. Vygotsky. 1978. *Mind In Society: The development of higher mental process.* Cambridge, MA: Harvard University Press.

[89] Greg Walsh, Elizabeth Foss, Jason Yip, and Allison Druin. 2013. FACIT PD: A Framework for Analysis and Creation of Intergenerational Techniques for Participatory Design. In *Proceedings of the SIGCHI Conference on Human Factors in Computing Systems* (CHI '13), 2893–2902. https://doi.org/10.1145/2470654.2481400

[90] Pamela Wisniewski, Haiyan Jia, Heng Xu, Mary Beth Rosson, and John M. Carroll. 2015. "Preventative" vs. "Reactive": How Parental Mediation Influences Teens' Social Media Privacy Behaviors. In *Proceedings of the 18th ACM Conference on Computer Supported Cooperative Work & Social Computing* (CSCW '15), 302–316. https://doi.org/10.1145/2675133.2675293

[91] Pamela Wisniewski, Heather Lipford, and David Wilson. 2012. Fighting for my space: coping mechanisms for sns boundary regulation. 609. https://doi.org/10.1145/2207676.2207761

[92]   Robert K. Yin. 2013. *Case study research: Design and methods.* Sage publications.

[93]   Jason C. Yip, Tamara Clegg, June Ahn, Judith Odili Uchidiuno, Elizabeth Bonsignore, Austin Beck, Daniel Pauw, and Kelly Mills. 2016. The Evolution of Engagements and Social Bonds During Child-Parent Co-design. In *Proceedings of the 2016 CHI Conference on Human Factors in Computing Systems* (CHI '16), 3607–3619. https://doi.org/10.1145/2858036.2858380